\newread\testifexists
\def\GetIfExists #1 {\immediate\openin\testifexists=#1
	\ifeof\testifexists\immediate\closein\testifexists\else
        \immediate\closein\testifexists\input #1\fi}
\def\epsffile#1{Figure: #1} 	

\immediate\openin\testifexists=e
	\ifeof\testifexists\immediate\closein\testifexists\else
        \immediate\closein\testifexists\input e\fipsf 
  
\magnification= \magstep1	
\tolerance=1600 
\parskip=5pt 
\baselineskip= 5 true mm \mathsurround=1pt
\font\smallrm=cmr8

\font\medrm=cmr9  
\font\medit=cmti9

\font\bigbf=cmbx12
 	\def\Bbb#1{\setbox0=\hbox{$\tt #1$}  \copy0\kern-\wd0\kern .1em\copy0} 
	\immediate\openin\testifexists=a
	\ifeof\testifexists\immediate\closein\testifexists\else
        \immediate\closein\testifexists\input a\fimssym.def 
\def\secbreak{\vskip12pt plus .5in \penalty-200\vskip 0pt plus -.4in} 
\def\hugeskip{\vskip12mm plus 3mm}
\def\Narrower{\par\narrower\noindent}	
\def\Endnarrower{\par\leftskip=0pt \rightskip=0pt} 
\def\br{\hfil\break}	\def\ra{\rightarrow}		
\def\a{\alpha}          \def\b{\beta}   \def\g{\gamma}  
          \def\D{\Delta}  \def\e{\varepsilon}
\def\h{\eta}            \def\k{\kappa}           \def\L{\Lambda} 
\def\m{\mu}             \def\f{\phi}                
\def\n{\nu}             \def\j{\psi}    
\def\r{\varrho}         \def\s{\sigma}  
\def\t{\tau}            \def\th{\theta}      
                     
\def\w{\omega}                            

 \def\LL{{\cal L}}

\def\cl{\centerline}    
\def\ni{\noindent}      \def\pa{\partial}       \def\dd{{\rm d}}        
\def\tl{\tilde}                 \def\bra{\langle}       \def\ket{\rangle}
 
\def\fn#1{\ifcase\noteno\def\fnchr{*}\or\def\fnchr{\dagger}\or\def
	\fnchr{\ddagger}\or\def\fnchr{\medrm\S}\or\def\fnchr{\|}\or\def
	\fnchr{\medrm\P}\fi\footnote{$^{\fnchr}$}{\scrunch#1\toe}\ifnum\noteno>4\global\advance\noteno by-6\fi
	\global\advance\noteno by 1}
 	\def\scrunch{\baselineskip=11 pt \medrm}
 	\def\toe{\vphantom{$p_\big($}}
	\newcount\noteno

\def\ffract#1#2{{\textstyle{#1\over#2}}}
\def\fract#1#2{\raise .35 em\hbox{$\scriptstyle#1$}\kern-.25em/
	\kern-.2em\lower .22 em \hbox{$\scriptstyle#2$}}

\def\half{\ffract12} \def\halff{\fract12}

\def\part#1#2{{\partial#1\over\partial#2}} 
 \def\ref#1{${\vphantom{)}}^#1$}

\def\bbf#1{\setbox0=\hbox{$#1$} \kern-.025em\copy0\kern-\wd0 
        \kern.05em\copy0\kern-\wd0 \kern-.025em\raise.0433em\box0}              

\def\eq{\ =\ }

\def\low#1{{\vphantom{]}}_{#1}} 
  
\def\ref#1{$\vphantom{.}^{\hbox{\smallrm #1}}$}

\def\Gbar{\raise.13em\hbox{--}\kern-.35em G}
\def\lap{\setbox0=\hbox{$<$}\,\raise .25em\copy0\kern-\wd0\lower.25em\hbox{$\sim$}\,}
\def\glt{\setbox0=\hbox{$>$}\,\raise .25em\copy0\kern-\wd0\lower.25em\hbox{$<$}\,}
\def\gap{\setbox0=\hbox{$>$}\,\raise .25em\copy0\kern-\wd0\lower.25em\hbox{$\sim$}\,}
   \def\newsect#1{\secbreak\noindent{\bf #1}\medskip}
\def\inst{{\rm inst}}\def\det{{\rm det\,}}
\def\GV{\hbox{ GeV}}\def\MV{\hbox{ MeV}}

{\ }\vglue 1truecm
\rightline{SPIN-1999/01}
\rightline{hep-th/9903189}
\hugeskip
\cl{\bigbf The Physics of Instantons}\medskip
\cl{\bigbf in the Pseudoscalar and Vector Meson Mixing}

\hugeskip

\cl{Gerard 't Hooft }
\bigskip
\cl{Institute for Theoretical Physics}
\cl{University of Utrecht, Princetonplein 5}
\cl{3584 CC Utrecht, the Netherlands}
\smallskip
\cl{and}
\smallskip
\cl{Spinoza Institute}
\cl{Postbox 80.195}
\cl{3508 TD Utrecht, the Netherlands}
\smallskip\cl{e-mail: \tt g.thooft@phys.uu.nl}
\cl{internet: \tt http://www.phys.uu.nl/\~{}thooft/	}
\hugeskip
\ni{\bf Abstract}\Narrower
	When the theory of Quantum Chromodynamics was introduced, 
it was to explain the observed phenomena of quark confinement
and scaling. It was then discovered that the emergence of 
instantons is an essential consequence of this theory. This
led to unanticipated explanations not only for the anomalously
high masses of the $\eta$ and the $\eta'$ particles, but also 
for the remarkable differences that have been observed 
in the mixing angles for the pseudoscalar mesons and the vector mesons.
\Endnarrower
\hugeskip\newsect{1. Introduction.}
	The discovery of ``The Eightfold Way"\ref1 in 1961 implied that
all observed mesons could be placed in $\,\bf 8\,$ or $\,\bf 1\,$ 
representations of the group  $SU(3)$, which later became\ref2 the flavor group 
$SU(3)^{\rm flavor}$, with the quarks $u$, $d$ and $s$ forming the fundamental $\,\bf 3\,$
representation. 
It was clear, however, that $SU(3)^{\rm flavor}$ is broken, and consequently,
mixing should take place between the eighth members of the octets
and singlet states. Later in the sixties, this became a hot
topic, when it appeared that this mixing for the pseudoscalar
mesons is very different from what happens with the vector mesons.
In terms of their quark components, we write the mesonic wave functions
$|\f_1\ket$ and $|\f_2\ket$ as
$$\eqalign{|\f_1\ket&=\cos\th\left({u\bar u+d\bar d-2s\bar s \over
\sqrt6}\right)+\sin\th\left({u\bar u+d\bar d+s\bar s \over\sqrt3}
\right)\,,\cr
|\f_2\ket&=-\sin\th\left({u\bar u+d\bar d-2s\bar s \over
\sqrt6}\right)+\cos\th\left({u\bar u+d\bar d+s\bar s \over\sqrt3}
\right)\,,}\eqno(1.1)$$

Great experimental efforts went into precisely determining these
mixing angles. Indeed, an experimental set-up, especially designed to
study the radiative decays of vector and pseudoscalar mesons,
was designed and built at CERN\ref3. Not only\ref4 the $(\w-\f)$ and the
$(\r-\w)$ mixings\ref5 were determined but also, by measuring its $2\g$
decay\ref{6}, a meson that at that time was called the $X^0$, could be
identified as being the ninth pseudoscalar meson; it was renamed
$\eta'$. 

The outcome of these measurements was indeed remarkable. In the
pseudoscalar case, the mixing angle turned out to be
$$\th_{PS}\approx 10^\circ\,,\eqno(1.2)$$
whereas the vector mesons mix with an angle
$$\th_V\approx 51^\circ\,.\eqno(1.3)$$
 
This gives
$$J^{PC}=0^{-+}\quad\Bigg\{\matrix{\eta(\approx 550 \MV)
&\displaystyle\approx
{.50\,(u\bar u+d\bar d)-.70\,(s\bar s)}\,,\vphantom{\big|_p}\cr 
\eta'(\approx 960 \MV)&\displaystyle\approx {.49\,(u\bar u+d\bar 
d)+.71\,(s\bar s)}\,,\vphantom{\big|^k}}\eqno(1.4)$$ and
$$J^{PC}=1^{--}\quad\Bigg\{\,\matrix{\w(\approx 780 \MV)
&\displaystyle\approx
{.71\,(u\bar u+d\bar d)-.06\,(s\bar s)}\,, \vphantom{\big|_[}\cr 
\f(\displaystyle\approx 1020 \MV)&\approx {.04\,(u\bar u+d\bar 
d)+1.00\,(s\bar s)}\,,\vphantom{\big|^k}}\eqno(1.5)$$

so we see that $\eta$ and $\eta'$ are divided to a large extent as dictated by $SU(3)$,
whereas $\w$ and $\f$ divide mainly in accordance with their quark composition.

These values for the mixing angles cannot be accidental, but should 
be explained. Many attempts were made to obtain some
insights. Then, in the early seventies, it was realized that meson
dynamics can be understood as being described by a non-Abelian
gauge theory with gauge group $SU(3)_c$ (the ``color group"), and 
fermions in the fundamental $\,\bf 3\,$ representation of this
group (the ``quarks")\ref7. Since the input parameters of
this theory, now called ``Quantum Chromodynamics" (QCD), 
appeared to consist only of the color gauge coupling parameter
$g$ (or, equivalently, the parameter $\L_{QCD}$) and the quark masses 
$m_f$, the mixing angles should, in principle,
be predictable. This, however, only added to the mystery: why are
pseudoscalar mesons so much different from the vectors?

Actually, there were other, even more distressing problems associated 
to the pseudoscalars. The successes of low-energy current algebra
considerations such as CVC (the conserved isovector vector current)\ref8 and PCAC
(the partially conserved isovector axial vector current)\ref9, strongly indicated that
meson physics has an approximate chiral $SU(2)\otimes SU(2)$ symmetry.
The pions, with their anomalously low mass values, can then be regarded
as being the Goldstone bosons associated with this symmetry.
It is easy to incorporate this symmetry in the QCD lagrangian, simply
by postulating that the $u$ and the $d$ quarks must have very small
mass terms here. The problem one then encounters is that, if this
were the case, QCD should actually have an even larger symmetry:
$U(2)\otimes U(2)$, which differs from the observed symmetries by an
extra chiral $U(1)$ component, and this should be reflected in a
(partially) conserved isoscalar axial vector current, $J^A_\m(x)$. Thus, the
symmetry held responsible for the relatively small value of the pion
masses, should necessarily induce another symmetry in the model that
would strongly reduce the mass of yet another particle: the pions
should have had a pseudoscalar partner, somewhat like the $\eta$, but
composed predominantly of $u\bar u$ and $d\bar d$ quarks, in the combination
$(u\bar u+d\bar d)/\sqrt2$ (which we will refer to as the state $\pi^o_o$). Adding the
strange quark $s$, should then only result in having an  extra
pseudoscalar meson made of  pure $s\bar s$, and, since its mass would be
constrained by the same terms that produce the kaon mass (the strange quark
mass term), the $s\bar s$ pseudoscalar meson could not be much heavier than
the kaon. It appeared that the kaon mass times $\sqrt2$, or $700\MV$,
should be an upper limit.\fn{This result can easily be seen from the 
equations for the masses in the appendix, by substituting   $\k=0$ and $M_\eta=M_\pi$ in Eqs~(A.4)--(A.6),
realizing that $B>0$ because of (A.9) and (A.10).
The limit is reached if $F_1\gg |F_3-F_1|$.}

In the early days, it was therefore suspected that QCD requires
explicit correction terms; after all, its ability to keep quarks 
permanently confined inside hadronic configurations was also not yet
explained.\ref{10}

Surprisingly, no such correction terms are needed at all. Both confinement
and the absence of the chiral $U(1)$ symmetry can now be adequately explained
as being special features of QCD alone. Both are due to special topological
aspects of the system. Confinement is due to the existence of color-magnetic
charges that undergo Bose condensation\ref{11}, and the absence of chiral $U(1)$
is due to instantons\ref{12}.

\newsect{2. Instantons.}

The first topological structures in gauge theories were the Abrikosov
vortices in superconducting material. They can be viewed as soliton
solutions in 2 space-dimensions. When particle physicists\ref{13} began
thinking of quarks being held together by stringlike structures, the
stringlike nature of these vortices caught their attention. It was realised
that all one needs is an {\it Abelian\/} Higgs theory\ref{14},
and the existence of such vortices is guaranteed.

When this vortex for the non-Abelian case was examined more closely,  it was found
to be unstable, and this implied the existence of other topologically
stable objects, but in 3 rather than 2 dimensions. These 3-dimensional
objects had to be magnetic monopoles\ref{15}. Bose condensation of color-magnetic
monopoles is now the favored explanation of quark confinement\ref{11}.

This subsequently raised another question: are there topologically stable objects
in more than 3 dimensions? What about 4 dimensions, and what would their
physical interpretation be? A localised object in 4 dimensions describes
an {\it event\/} rather than a particle, and so we devised the name
``instanton" for such objects\ref{16}. The first example in gauge theory
had been described by Belavin et al\ref{17} in 1975. In their paper, Minkowski
space had been replaced by Euclidean space. In this space, they found localised 
solutions of the classical gauge field equations. This raised questions such as:
what kind of events do these
instantons correspond to, and why do they exist only in Euclidean space?

Euclidean spacetime is obtained upon analytic continuation of physical
amplitudes for imaginary time: we replace $t$ by $i\,x_4$ with now $x_4$
a real coordinate. This is exactly what one needs to do if one wishes to
compute a {\it tunnelling amplitude}, replacing the usual perturbation
expansion by a BKW expansion. The exponential suppression factor in the amplitude 
is obtained by solving the classical equations with time being replaced by an
imaginary parameter. Therefore, instantons are to be interpreted as tunnelling
events. Indeed, their contributions to physical amplitudes are proportional
to $\exp(-8\pi^2/g^2)$, an exponential suppression typical for tunnelling. 
But this is not all. The tunnelling event in question {\it violates a
conservation law\/} that would be respected by any ordinary perturbative
effect. Which conservation law? Belavin et al had noted in passing that their
solution has
$$\int\dd^4x\, F^a_{\m\n}\tl F^a_{\m\n}=\pm{32\pi^2\over g^2}\,,\eqno(2.1)$$
where $\tl F_{\m\n}=\half\e_{\m\n\a\b}F_{\a\b}$, and the sign refers to instantons
and anti-instantons, respectively. But, according to  
Adler\ref{18}, and Bell and Jackiw\ref{19},
$$\pa_\m J^A_\m={g^2\over 16\pi^2}F^a_{\m\n}\tl F^a_{\m\n}\,.\eqno(2.2)$$
This is the well known axial triangle diagram anomaly. One-loop enormalisation effects cause
an apparently tiny (proportional to $g^2$) violation of axial current
conservation. We observe that the instanton would give rise to a transition between states
with different values for the axial charge $Q_5(t)=\int\dd^3x\,J_0^A({\bf x},t)$:
$$\D Q_5=Q_5(T)-Q_5(-T)=\int_{-T}^{T}\dd t\int\dd^3 {\bf x}\ \pa_0 J^A_0(x)=\pm 2\,.\eqno(2.3)$$
Note that, although the tunnelling amplitude is computed using analytic continuation
to Euclidean time, the actual event takes place in Minkowski space-time. Being
topological equations, Eqs.~(2.1) --- (2.3) hold {\it both\/} in Euclidean and in Minkowski
space-time, since they are independent of the metric $g_{\m\n}$. This is why it is permitted
to use the real charge density $J_0({\bf x},t)$ in Eq.~(2.3).

According to a theorem by Adler\ref{18} and Bardeen\ref{20}, the anomaly equation~(2.2) is
essentially not affected by any renormalisation beyond the one-loop level. 
Apparently, the number 2 in Eq.~(2.3)
will not be affected by higher order corrections.\fn{Nor is there need to worry
about the fact that $g$ is a running coupling strength. In fact, the fields
$F_{\m\n}$ in Eqs.~(2.1) and (2.2) should be replaced by ${\cal F}_{\m\n}=g F_{\m\n}$,
so that $g$ no longer appears explicitly.}\ The meaning of this is clear.
One left-handed polarised quark (contributing $+1$ unit to the axial charge)
is turned into a right-handed one (with $Q_5=-1$), or vice versa. It is
important to note that in a theory with $N_f$ quark flavors, Eq.~(2.3) holds
for each flavor separately. In total, one therefore has
$$\D Q_5=\pm 2N_f\,.\eqno(2.4)$$

There are several ways to understand, mathematically as well physically, why
and how such transitions take place\ref{12}. Here we will limit ourselves
to the explanation that, in a properly regularised and renormalised theory, 
the total number
of Dirac levels for fermions, in a given volume, is precisely specified.
The instanton causes exactly one such level to make a transition from 
positive to negative energy,
or vice-versa, thus crossing the Fermi level of the vacuum. This way,
one quark with one helicity may be materialised from the Dirac sea, while
another, with opposite helicity, submerges into this sea. More precise
and complete explanations have been given elsewhere\ref{12, 21}.

Not only instantons violate chiral $U(1)$ invariance, but also the quark
mass terms. This implies that a new phase angle, $\th_\inst$, emerges in the
description of interference between these two symmetry breaking effects. 
In the early days
of QCD, it had not been realized that QCD possesses {\it two\/} fundamental parameters,
the gauge coupling $g$ and the instanton angle $\th_\inst$.
In the QCD lagrangian, the effect of this angle can be described by adding a
term 
$$i\th_\inst\cdot{g^2\over 32\pi^2}F_{\m\n}\tl F_{\m\n}\,.\eqno(2.5)$$
In perturbation expansion, this term seems to give no effect at all because it
can be written as a pure derivative:
$$F_{\m\n}^a\tl F_{\m\n}^a={16\pi^2\over g^2}\pa_\m K_\m=\pa_\m\big(2\e_{\m\n\a\b}A_\n^a(\pa_\a 
A_\b^a+{g\over3}f^{abc} A_\a^b A_\b^c)\big)\,.\eqno(2.6)$$
$K_\m$ is the Chern-Simons current. Because of this equation, all Feynman
diagrams with this vertex in them carry a factor $\sum_ip_\m^{(i)}$, the sum
of all external momenta, and hence they vanish. Instantons nevertheless produce 
non-trivial physical effects depending on $\th_\inst$, only because $K_\m$ is
not gauge-invariant. Since instantons are the only stable objects with a
non-vanishing value of the integral (2.1),  they are the only structures
that can yield $\th_\inst$-dependent effects.

\newsect{3. Instantons in QCD.}
Thus, instantons produce a new kind of interaction in all non-Abelian gauge 
theories, and in particular in QCD. It is known since 1976 that this
interaction can be mimicked by an `effective interaction lagrangian' of the
form\ref{16}
$$\LL^\inst(x)= \k\, e^{i\th_\inst}\det\{-\bar\j_R(x)\j_L(x)\}\ +\ \hbox{h.c.}\,.\eqno(3.1)$$ 
Here, $\k$ is a constant that should be in principle computable, and it
contains the factor $e^{-8\pi^2/g^2}$. The subscripts $L$ and $R$ refer to the
left- and right handed helicities, obtained by means of the projection
operators $\half(1\pm\g^5)$. The determinant is the determinant
of the matrix $\bar\j^b_R\j^a_L$, where $a$ and $b$ are flavor indices only.
The color indices and the Dirac spin indices can be arranged in 
several ways (this can be computed explicitly\ref{16}), but for simplicity we will 
ignore these, since this effective
lagrangian must be seen as active in an effective hadron model, and we limit ourselves
to colorless scalar and pseudoscalar mesons. Note that, in any case, vector mesons
consist of quark-antiquark pairs that are either both left-handed or both right handed, so that 
the determinant (3.1) will have no effect on them (apart from higher orders). 
We see that the interaction (3.1) has exactly the right
quantum numbers for absorbing $N_{\rm flavor}$ left helicity fermions and
creating an equal number of right handed ones (or vice-versa). In particular,
the determinant is easily seen to be the simplest possible interaction
that conserves $SU(N_f)\otimes SU(N_f)$ symmetry, while breaking $U(N_f)
\otimes U(N_f)$.

The cases $N_f=0$ and $N_f=1$ are rather special. If $N_f=0$ while $\th_\inst\ne 0$, 
the interaction (2.5), which is even under charge conjugation\fn{The charge conjugation operator
$C$ replaces the gluon field $A_\m$ by $-A_\m^*$, which is a non-trivial transformation already
in the pure $SU(3)$ gluon theory; for a purely gluonic $SU(2)_c$, $N_F=0$  theory,
$C$ would be indistinguishable
from a gauge transformation, and therefore trivial.}, but odd under parity, 
implies an explicit $P$ and $CP$ violation. One can show\ref{22} that the color-magnetic
monopoles obtain fractional electric charges, proportional to $\th_\inst$. This
should have physically observable effects. This is not a purely academic
statement, because at very low energies, in QCD, one may regard the up and down
quarks as being heavy. If $N_f$ were equal to one, the interaction (3.1) would
blend with the quark mass term. In this case, no symmetry would protect the
single quark flavor from getting a mass induced by QCD interactions.

Of particular interest are QCD models with $N_f\ge 2$. In this case, there is
a global chiral $SU(N_f)\otimes SU(N_f)$ symmetry that is not affected by instantons. In all such cases,
the effective interaction (3.1) would be non-renormalisable. This means that
no {\it perturbative\/} interaction of this sort may be admitted. The (non-perturbative) interaction
shows up only at low energies. Indeed, at higher energies, one must substitute
the running value of $g^2$ in $\k$, so that the effective strength of this coupling
rapidly decreases with energy.

Studying the case $N_f=2$ gives us the physics of QCD if we allow ourselves to
neglect the effects of the strange quarks.
In Ref\ref{12}, a low energy effective meson model for QCD with instanton effects
included is discussed at length. Here, we summarise its results. The effective
meson fields $\f_{ij}$ basically correspond to the composite operators
$\bar q_{Rj}q_{Li}$, and this $2\times2$ matrix is decomposed into eight real
mesonic fields: a scalar isoscalar $\s$, a pseudoscalar isoscalar $\eta$, a scalar isovecor $\vec\a$, and a
pseudoscalar isovector $\vec\pi$. We write
$$\f=\half(\s+i\eta)+\half(\vec\a +i\vec\pi)\cdot\vec\t\,,\eqno(3.2)$$
where $\t^{1,2,3}$ are the Pauli matrices. The interaction (3.1) now looks as\fn{A few signs
here are chosen to be different from Ref\ref{12}.} 
$$\eqalign{&\LL^\inst= U+U^\dagger\ ,\cr U\ =\ \k\,e^{i\th_\inst}\det(-\bar q_Rq_L)\ =\ \k&\,e^{i\th_\inst}\det\f\ 
=\ \k\,e^{i\th_\inst}\big((\s+i\eta)^2-(\vec\a+i{\vec\pi})^2\big)\,. }\eqno(3.3)$$
Here, the parameter $\k$ differs from the $\k$ in (3.1) by some coefficient. This is because,
in (3.1), the $\j$ fields were defined such that the kinetic terms in the Lagrangian are
normalised to $\bar\j(\g D+m_f)\j$, whereas in (3.3), we assume the kinetic terms for
the mesons to be normalised to ${\rm Tr}\,\pa_\m\f\pa_\m\f^\dagger=\half\big((\pa_\m \s)^2+(\pa_\m
 \eta)^2+(\pa_\m \vec\a)^2+(\pa_\m \vec\pi)^2\big)$.
 
In Ref\ref{12}, it is explained that if the $u$ and the $d$ quark masses may be neglected
then $\th_\inst$ will be aligned to zero, and the effective coupling goes as
$$\LL^\inst\ra 2\k(\s^2+\vec\pi^2-\eta^2-\vec\a^2)\,.\eqno(3.4)$$
Now, since this is the only effect that splits the pion from the eta, and since
the pion continues to behave as a massless Goldstone boson, one can deduce from
(3.4) that the eta mass becomes
$$m^2_\eta=8\k\eqno(3.5)$$
(both $\vec\a$ and $\s$ were already massive before the instantons were switched on, 
because of $U(2)\times U(2)$ invariant potential terms in the unperturbed Lagrangian,
see Ref\ref{12}).
The beauty of this simple analysis is that the instanton interaction bares 
exactly the quantum numbers required for the eta mass term\fn{Remember that we are still
discussing the two-flavor case, so the effects of the strange quarks are ignored. Therefore,
eta here stands for the pseudoscalar state $\pi^0_0 = \ffract1{\sqrt2}(u\bar u+d\bar d)$.}.
Continuing the
analysis furthermore shows that the operator $F_{\m\n}\tl F_{\m\n}$ has the same
quantum numbers as the eta field, and so, one expects a considerable mixture
between the eta and pure gluonic matter.\fn{If one adds the third flavor $s$ here, this
object will become a flavor octet instead of a singlet and hence not mix with glue but 
predominantly with $s\bar s$, whereas the privilege to mix with substantial amounts of pure glue
will be reserved for the ninth pseudoscalar meson, $\eta'$.}

Although, in principle, not only the pseudoscalars, but also the vector mesons could 
mix with gluonic matter, such
a vector meson mixing is not directly associated to instantons, as mentioned when discussing the 
effective instanton action (3.1). In Sect.~5, we exlain why
it is much weaker than for the pseudoscalars.

The above, however, still ignored the presence of strange quarks. Adding the strange
quark gives the effective interaction dimension 3 in terms of the mesonic fields:
$$\det(\f)=\f_{11}\f_{22}\f_{33}\pm\cdots\,.\eqno(3.6)$$
This does not have the quantum numbers of the mass terms of either the $\eta$ or 
the $\eta'$ particles.
Here, it is necessary to consider the consequences of chiral $U(3)\otimes U(3)$
breaking more carefully. As the $SU(3)$ case was not discussed in detail in Ref\ref{12},
we give a short review here. 
\def\Tr{{\rm Tr\,}}\def\Re{{\rm Re\,}}

\newsect{4. A discussion of the 3 flavor case.}
It is instructive to describe the scalar and the pseudoscalar mesons in terms of a
simple model. The model of Ref\ref{12} can extended to the $N_f=3$ case
without any major changes. The meson fields are written in the form of a matrix field
$\f_{ij}$ which, as before, is assumed to have the quantum numbers of the quark-antiquark composite
operator\fn{The minus sign was chosen here so as to achieve $\bra\phi\ket>0$ while keeping
the sign convention for the mass terms of Eq.~(4.4).}
$-\,\bar q_{Rj}\,q_{Li}$. Under a chiral $U_L\otimes U_R$ transformation, it transforms
as $$\f_{ij}'=U_{ik}^L\f\low{k\ell}U_{\ell j}^{R\dagger}\,.\eqno(4.1)$$
The lagrangian is taken to be
$$\LL=-\Tr\pa_\m\f\pa_\m\f^\dagger-V(\f)\,,\eqno(4.2)$$
where 
$$\eqalignno{V(\f)&\eq V_0+V_m+V_\inst\,;&\cr
V_0&\eq -\m^2\Tr\f^\dagger\,\f+A(\Tr\f^\dagger\,\f)^2+B\,\Tr(\f\,\f^\dagger\,\f\,\f^\dagger)
\,,&(4.3)\cr
V_m&\eq -2m_u\Re\f_{11}-2m_d\Re\f_{22}-2m_s\Re\f_{33}\,,&(4.4)\cr
V_\inst&\eq-2\k\,\Re(e^{i\th_\inst}\det\f)\,.&(4.5)}$$
Here, $V_0$ has the complete $U(3)\otimes U(3)$ symmetry; $V_m$ represents the contributions
of the quark mass terms, breaking the symmetry down to $U(2)\otimes U(2)$ if $m_u$ and $m_d$
are small (note that, here, the parameters $m_{u,d,s}$ are proportional to the current quark mass
terms, but they do not carry the dimensions of a mass). $V_\inst$ represents the 
instanton contribution, which, having the form of a determinant, breaks the symmetry
into $SU(3)\otimes SU(3)\otimes U(1)^{\rm vector}$. The coefficient $\k$ contains the
standard exponential term\fn{Note that, in this term, $g$ is a running coupling strenth. In an accurate
analysis\ref{16}, this matches the non-trivial canonical dimension of this interaction. Therefore,
the exponential coefficient ends up to be of order one in units $\L_{QCD}$.}\  $\exp(-8\pi^2/g^2)$.

The signs in the definition of $\f_{ij}$ and in Eqs~(4.3)--(4.5), were chosen in such a
way that the vacuum expectation values will be positive. $\m$, $A$ and $B$ are parameters
of the model that must obey 
$$\m^2>0\ ,\qquad A+B>0\ ,\qquad 3A+B>0\,.$$ We take
$$\f=\pmatrix{F_1&0&0\cr 0&F_2&0\cr0&0&F_3}+\tl\f\,,\eqno(4.6)$$
where $F_i$ are the vacuum expectation values. Imposing that the terms linear in
the quantum fields $\tl\f$ cancel out, gives us the equations for the $F_i$.
If $\th_\inst\ne0$, these numbers in general will be complex, and mixture occurs
between the scalars and the pseudoscalars, which makes the computations very
lengthy. In experimental observations the value of $\th_\inst$ is found to be very close to zero,
or in other words, there is no observed mixing between scalars and pseudoscalars, since otherwise there would have been
substantial parity violation in the strong interactions.
For simplicity, we will therefore now take $\th_\inst$ to be zero. Redefining
$$R=2A(F_1^2+F_2^2+F_3^2)-\m^2\,,\eqno(4.7)$$ we get:
$$R+2B\,F_1^2={m_u+\k\,F_2F_3 \over F_1}\ , \eqno(4.8)$$
and its permutations, replacing $m_u$ by $m_d$ and $m_s$.

Just as in Eq.~(3.2), the real components are scalar fields, and the
imaginary parts are pseudoscalars. We will denote the scalars by $S$ and the
pseudoscalars as $P$.

It is now worth-while to compute the masses and mixing angles in this
model. Writing \def\pl{\ +\ }
$$\tl\f =\pmatrix{S_1&S_{12}&S_{13}\cr S_{12}^*&S_2&S_{23}\cr S_{13}^*&S_{23}^*&S_3 }
+i\pmatrix{P_1&P_{12}&P_{13}\cr P_{12}^*&P_2&P_{23}\cr P_{13}^*&P_{23}^*&P_3 }\,,\eqno(4.9)$$
we expand the potential $V(\f)$ up to the terms quadratic in $S$ or $P$:
$$\eqalign{V(\f) \eq V(F) \pl S_1^2\quad&\big(R+(4A+6B)F_1^2\big)\cr
\pl 2S_1S_2\ &\big(4A\,F_1F_2-\k F_3\big)\cr
 \pl 2|S_{12}|^2&\big(R+2B\,F_1 F_2+\k F_3+2B(F_1^2+F_2^2)\big) \cr
 \pl P_1^2\quad&\big(R+2B\,F_1^2\big)\cr\pl  2P_1 P_2\ &(\k F_3) \cr 
 \pl 2|P_{12}|^2&\big(R-2B\,F_1 F_2-\k F_3+2B(F_1^2+F_2^2)\big)\cr
 \pl\quad&\hbox {the two cyclic permutations}\ .}\eqno(4.10)$$

Using Eq.~(4.8), the part depending on the fields $P_1$, $P_2$ and $P_3$ (the neutral
pseudoscalars) can be simplified into
$$\k F_1F_2F_3\Big({P_1\over F_1}+{P_2\over F_2}+{P_3\over F_3}\Big)^2\pl
{m_u\over F_1}P_1^2+{m_d\over F_2}P_2^2+{m_s\over F_3}P_3^2\,,\eqno(4.11)$$
which shows that, in the chiral limit ($m_i=0$), where all $F_i$ are equal,
only the ninth component of the pseudoscalars gets a mass, which is proportional
to the instanton coefficient $\k$. The scalar mesons $S$ will always have masses
due to the regular interactions (4.3).

Substituting the angles that have been measured, Eqs~(1.4) and (1.5), and the
masses of $\pi^0$, $\eta$ and $\eta'$, gives us the numbers $\k F_i$ and
$m_i/F_i$:\def\GV{\hbox{ GeV}}\def\MV{\hbox{ MeV}}
$$\eqalign{\k F_1\approx \k F_2 \approx .22\GV^2\ ,\qquad& \k F_3\approx .28\GV^2\,;\cr
m_u/F_1+m_d/F_2=2M_\pi^2=.0365\GV^2\ ,\qquad&m_s/F_3=.44\GV^2\,.}\eqno(4.12)$$
This gives the ratio
$${2m_s\over m_u+m_d}\approx 30.7\,.\eqno(4.13)$$

The $\pi^\pm$ and the $K$ masses are now computable. Since the input parameters
had an exact isospin invariance, $\pi^\pm$ are degenerate with the $\pi^0$,
and $K^\pm$ with $K^0$. The  $K$ mass-squared corresponds to the coefficient
in front of $|P_{13}|^2$ and $|P_{23}|^2$ in (4.10), which is
$$M_K^2=R+2B(F_1^2+F_3^2-F_1F_3)-\k F_2\,,\eqno(4.14)$$ and since all numbers in
here were already determined by the $\eta$ and $\eta'$ masses and mixing angles,
the outcome is `predicted', yielding
$$M_K\approx 509\MV\,,\eqno(4.15)$$
no more than 3\% away from the actual value. Although this beautiful agreement with the
experimental value of the $K$ mass may be accidental, this does indicate that the
mechanism for chiral symmetry breaking described here is realistic. In fact, this just
confirms the long known fact that the meson masses squared are approximately linearly proportional
to the quark masses.

It is probably even better to use the observed value for the kaon mass, $495\MV$, to 
estimate the strange quark mass term, $m_s$. One then gets, instead of (4.13):
$${2m_s\over m_u+m_d}={M_K^2\over M_\pi^2}\Big(1+{F_3\over F_1}\Big)-1
\approx 29\,.\eqno(4.16)$$

The scalar mesons in this model are fairly heavy. We found the scalar pion
to be in the range $1340$ to $1580\MV$, and the scalar kaon to be about $150\MV$ heavier
than the scalar pions. The masses and mixing angles of the
scalar counterparts of $\eta$ and $\eta'$, here called $\s$ and $\s'$, depend 
explicitly on the parameter $A$, which was
not yet determined (see the Appendix for further details). We observe, 
that our rather crude model of Eqs.~(4.2)--(4.5) gives a
quite realistic phenomenology for the pseudoscalar mesons. 

\newsect{5. A few words on $\w$--$\f$ mixing and the $\eta_c$.}

The model of Sect.~4 is to be regarded as a {\it low energy, effective\/} theory.
The scalar resonances predicted in the region of $1340$ to $1580\MV$ are expected to be quite wide. 
Their mass formulas are given in the Appendix. From the way the masses depend on the parameters
$A$ and $B$ one deduces that the scalar masses are very much model dependent. Indeed,
an alternative model can be constructed in which the fields $\f$ obey a non-linear unitarity
constraint: $\f\f^\dagger=\Bbb I$. This model only contains pseudoscalars; the scalar masses
were sent to infinity.

At high energies, QCD is more effectively described in terms
of vortex dynamics; at higher energies still, the asymptotic perturbation expansion
of asymptotically free QCD is the best. The question why $\w$ and $\f$ only mix rather weakly,
and have nearly no gluon content, can be explained in the high-energy limit.
Replacing these particles by the $J/\j$, we can make the following observation.

Whereas $\eta_c$ couples to gluonic states via intermediate states with only two
gluons, it is well-known that $J/\j$ needs a 3-gluon intermediate state to decay.
This means that, in the limit $m_c\ra\infty$, $J/\j$ is coupled to gluonic matter
much more weakly than $\eta_c$. Intermediate gluonic states are the only way in
which a $c\bar c$ bound state can couple to other flavor states such as $u\bar u$
and $d\bar d$. We see that $J/\j$ is shielded from these other states by a factor
$\a_{\rm strong}$ relative to $\eta_c$. Clearly then, $J/\j$ will not hardly mix as strongly to these
other states as $\eta_c$ will do. If we now replace the charmed quark by the
strange quark, we may expect the same qualitative behaviour, although the precise 
numerical coefficients will be much harder to calculate. In any case, we should
not be surprised to find that the vector state $s\bar s$ hardly mixes with
$u\bar u$ and $d\bar d$; the reason for this is that the mixing goes via an
intermediate state of pure glue, and the coupling between the vector states and pure
glue is suppressed as compared to the pseudoscalar particles.

This argument must be added to the observation made at the beginning of Sect.~3, 
that the operator associated with the creation and absorption of vector mesons,
$\overline\j \g_\m\j$, contains either only left handed fermion-antifermion pairs or only right handed
ones, and therefore it does not match the quantum numbers of an effective instanton
interaction (as both the scalar and the pseudoscalar mesons do).

\midinsert\cl{\epsffile{instfig.ps}}
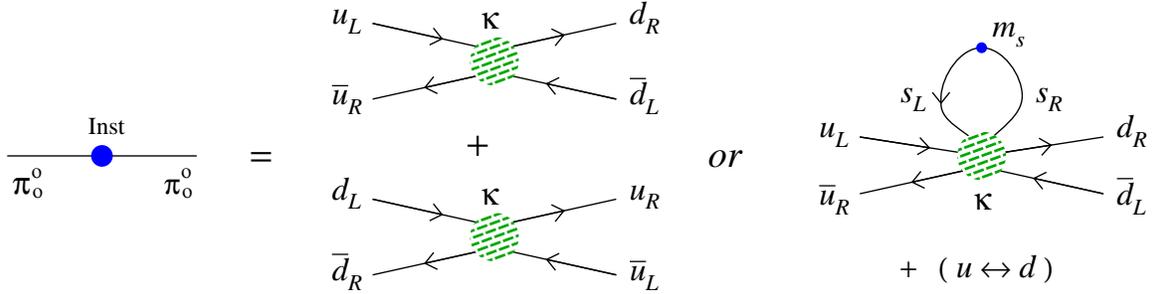
\Narrower {\scrunch Fig.~1. In the 2-flavor case, the quantum numbers of the effective instanton interaction
exactly match the quantum numbers of a mass term for the quark combination
$\pi^o_o\approx u\bar u+d\bar d$. In the 3-flavor case, there is an extra
{\medit s\/}-quark emitted and absorbed, with opposite helicities. This gives a contribution to
the $ \pi^o_o$ mass that is proportional to the strange quark mass. Then, because of the vacuum shift
described by Eq.~(4.6), this object mixes with $s\bar s$ and it becomes the physical $\eta$
particle, which is close to the $SU(3)^{\rm flavor}$ octet state 
$\h_8=(1/\sqrt6)(u\bar u+d\bar d-2s\bar s)$. The mass of the $ 
\eta'\approx(1/\sqrt3)(u\bar u+d\bar d+s\bar s)$ is not limited by the strange quark mass, 
but arises as described in Sect.~4; i.e., its mass is proportional to the instanton action
$\k$.\Endnarrower
}\endinsert

We now see that, in contrast, instantons give a fairly effective mixing
between all diagonal pseudoscalar states. Indeed, the model of Sect.~4 could be
used to study the charmed sector, in particular in an approximation where we
ignore the strange quark.

An alternative way to understand the two-flavor model of Sect.~3, and Ref\ref{12},
is to integrate first over all virtual strange-quark loops. A Feynman diagram
containing the interaction (3.1), can then be seen to yield an amplitude
proportional to a $U(2)$ determinant that does not contain the strange quarks, see Fig.~1. 
In order to yield a non-vanishing contribution, the strange quark, in its closed
loop, must switch its helicity, but this can happen due to the non-negligible
value of the strange quark mass. Indeed, the effective instanton interaction
is now proportional to $m_s$. The case for more heavy flavors is a bit more subtle.
Primarily, the effect of an instanton will carry as a factor the {\it product
of all "heavy" flavor masses}, $m_s\cdot m_c\cdot m_b\cdots$, but when they get
heavier than $\L_{QCD}$ the heavy flavors decouple, and the instanton behaves as
if they were not there. This is why effects due to charm, bottom and top are usually
not considered. These heavy flavors do not mix very much with the light ones,
be it via instantons or other forms of glue.

\newsect{6. Instantons and spontaneous chiral symmetry breaking.}

There are various other aspects of QCD dynamics that are directly associated to instantons.
One of these is the nature of the dynamical forces that cause the {\it spontaneous\/}
breakdown of chiral $SU(N_F)\otimes SU(N_F)$ down to the vector flavor symmetry group 
$SU(N_F)$. Why should such a spontaneous symmetry breaking occur at all? 

Several hand-waving arguments can be brought forward. We see spontaneous symmetry
breaking happen explicitly in some model calculations. In $1+1$ dimensions, QCD can be
solved exactly in the $N_c\ra\infty$ limit\ref{23}. In this limit, we see the Goldstone pions
emerge in the exact solution. They are also observed when QCD is solved on a lattice in
the large coupling limit\ref{24}. 

We would like to know whether QCD related theories can be constructed in which chiral
symmetry is {\it not\/} spontaneously broken, but realized explicitly in the Wigner mode.
Such a theory could be employed to describe a new strong interaction regime for the
weak interactions at ultra-high energies. It was attempted to construct a theory in 
which the presently elementary leptons and quarks are seen as bound states of a new
kind of quarks at ultra-high energies. An ultra-strong color force should bind these
new quarks. Such a theory, often referred to as ``Technicolor"\ref{25}, however 
only works if some symmetry protects the ordinary
quarks and leptons against developing too large mass terms. This symmetry can only be a
chiral symmetry that is {\it not\/} spontaneously broken. The attempts at constructing technicolor
theories for the electro-weak forces were unsuccessful. The demise of these theories
was partly due to the following insight concerning QCD related theories.

Using a background of classical flavor gauge fields, one can
derive the anomalies in the vacuum-to-vacuum amplitudes in given channels of these
classical  background fields\ref{26}. One does this by constructing instantons out of these background fields,
and asking how many axial charges are generated by it. The QCD Lagrangian gives us 
explicit and exact answers. We derive which of the chiral charges associated with the background
gauge fields are created or destroyed by these instantons, via the generalized version 
of Eq.~(2.4). Now consider some effective theory
describing the mesonic and baryonic bound states. Any effective model for these hadrons should reproduce the
{\it same\/} answers, i.e., the same chiral charges $Q_5$, with the same quantum numbers, should be
produced by our background instanton. This observation provides us with very strict constraints on the
spectrum of states that should be introduced in the effective meson model. If now the
chiral flavor symmetry is not spontaneously broken, the total amount of axial flavor charges
involved in the anomaly, as dictated by Eq.~(2.4), must be reproduced by the fermionic
(mesonic and baryonic) objects described in the effective model. This is called the `anomaly
matching condition'\ref{26}.

Explicit calculations then lead to a surprise: if the symmetry were realized in the Wigner
mode, one finds that, more often than not, the bound states 
carry too large axial charges to allow us to match the anomalies using
Eq.~(2.4). This would force us to consider models in which the `number of mesonic or
baryonic species' is fractional. Numbers such as 1/9 and 1/25 emerge in $SU(3)^{\rm color}$ and $SU(5)^{\rm color}$
theories, which would be an absurdity. If however the effective degrees of freedom are assumed
to realize the external symmetries in the Goldstone mode, then the constraints posed
by the anomaly matching condition can always be realized.
Thus we arrive at a contradiction if we assume the chirally symmetric Wigner mode to be
realized in QCD. An exception is QCD with exactly two flavors. It is the strange quark that
causes the first real problem in constructing a chirally symmetric spectrum. This leads
one to conclude that, if chiral symmetry were not already spontaneously broken in the
up + down sector of the flavor group, surely the strange quark would trigger spontaneous
chiral symmetry breakdown.

Thus the anomaly matching requirement rules out many attempts to use a QCD related theory
at the TeV scale and assume it to realize its chiral symmetries in some unconventional
way

\newsect{7. Conclusion and remarks.}
A simple polynomial lagrangian for the effective interactions between scalar
and pseudoscalar mesons, with in addition the simplest polynomial that reflects
the correct quantum numbers of an instantonic interaction, can reproduce the
observed meson spectrum quite reasonably. The fact that the pseudoscalars tend
to mix along the dividing lines of the $SU(3)$ octet and singlet representations, 
while the vectors $\w$ and $\f$ 
tend to mix in such a way that pure flavor bound states emerge, can be 
understood quite naturally. The pseudoscalar mesons can be addressed in a
simple model. In addition to pseudoscalar mesons, this model contains also scalar mesons, but no
vectors. If one wishes to include vector mesons, one has to turn the effective
model into an  $SU(3)^{\rm flavor}$ non-Abelian, spontaneously broken
gauge theory. Such a model contains much more mesonic fields and
freely adustable parameters, and consequently it gives little further insight.

The experimental evidence that there is little PC violation
in QCD indicates that $\th_\inst$ must be very small or zero. The sign of the instanton
interactions (which is a free parameter since one may freely choose $\th_\inst$), is
as indicated in the effective action term (3.1) (note the minus signs in Eqs.~(4.2),
(4.5) and in the definition $\f=-\bar q_R q_L$), with $\th_\inst\approx0$.

Within the paradigm of QCD, in the absence of weak interactions, it is not unnatural
to put $\th_\inst=0$, since only weak
interactions, with their explicit CP violation effects by having a phase angle in the Kobayashi-Maskawa
matrix, can send $\th_\inst$ away from zero.
Note however that, by rotating the field $\f$, one can transport the $\th_\inst$ angle from the
instanton term (4.5) to one of the quark mass terms in Eq.~(4.4). Since $m_u$ is the
tiniest mass term, it is most natural to put  $\th_\inst$ as a phase in the $u$ quark
mass term, which then becomes Re$(e^{-i\th_\inst}m_u\f_{11})$. As $m_u$ runs towards
smaller values at very high energy scales, its phase will be affected by the weak
interactions. As the TeV scale is reached, it becomes difficult to see why such
effects should stay extremely small. From the observed absence of scalar-pseudoscalar
mixing one must deduce however that the $\th_\inst$ is extremely small. How this
fine-tuning can be explained is as yet an unresolved problem. We suspect that
new physics at the TeV scale must be responsible.

The question whether instantons also lead to observable effects at high energies
is more difficult to answer. At high energies, where the running coupling
parameter $g_{\rm strong}$ tends to become small, instantons are very efficiently screened, as
their amplitudes vary as $\exp(-8\pi^2/g_{\rm strong}^2)$, so we do not expect
direct instanton effects at high energy. Instantons are very soft objects.

When the fields $\f$ are coupled to the electro-weak gauge field via their
currents (defined by their weak transformation rules), we hit upon a serious shortcoming
of this simple model: it does not reproduce the experimentally well-confirmed
$\D I=\halff$ rule. Indeed, neutral kaons tend to decay only into charged pions;
the $2\pi^0$ decay is suppressed! This is because the GIM mechanism prevents
the decay through neutral vector bosons. This deficiency impedes attempts to
investigate the $\e'/\e$ problem using models of this sort. To reproduce the
$\D I=\halff$ rule one must take renormalization group effects in QCD into account.

\newsect{Appendix A. Mass formulae for scalar and pseudoscalar mesons.}

The model of Sect.~4 appears to generate fairly decent estimates for mesonic
mass relations. We here give some of the formulae. The derivations are
straightforward. We start from the effective action described by Eqs.~(4.2) -- (4.5).
The masses are determined by Eqs.~(4.10). Because of isospin symmetry, and because
electromagnetism was ignored, we keep $m_u=m_d$, a number that in more refined
theories should be replaced by $\half(m_u+m_d)$. The two independent vacuum 
expectation values $F_1$ and $F_3$ are determined by Eq.~(4.8), which can be written
as
$$\eqalignno{F_1\,\big((4A+2B)\,F_1^2+2A\,F_3^2-\m^2-\k\,F_3\big)&=m_u\,,&(A.1)\cr
F_3\big(4A\,F_1^2+(2A+2B)\,F_3^2-\m^2\big)-\k\,F_1^2&=m_s\,.&(A.2)}$$
The pseudoscalar masses (after some algebraic manipulations) are then described by 
$$\eqalignno{M_\pi^2&= m_u/ F_1\ ;&(A.3)\cr M_K^2=M_\pi^2+(2B\,F_3+\k)(F_3-F_1)
&={m_s+m_u\over F_1+F_3}\ ;&(A.4)\cr
M_\eta^2+M_{\eta'}^2-2M_\pi^2&=2B(F_3^2-F_1^2)+3\k\,F_3\ ;&(A.5)\cr
{(M_\eta^2-M_\pi^2)(M_{\eta'}^2-M_\pi^2)\over M_K^2-M_\pi^2 }&=2\k\,(F_1+F_3)\ ;&(A.6)\cr
(M_{\eta'}^2-M_\eta^2)\sin 2\th_{PS}&={\textstyle{ 2\sqrt2\over3}}\big(2B(F_3^2-F_1^2)-\k\,(F_3-F_1)\big)\ .&(A.7) \cr
(M_{\eta'}^2-M_\eta^2)\cos 2\th_{PS}&=\ffract13\big(8\k\,F_1+\k\,F_3+2B(F_1^2-F_3^2) \big)\ .&(A.8) }$$
 
For completeness, we list here the formulae for the scalar mesons $\pi_S$, $K_S$, $\s$ and $\s'$, which are like
$\pi$, $K$, $\eta$ and $\eta'$ but with $J^{PC}=0^{++}$:
$$\eqalignno{M^2_{\pi S}&=M_\pi^2+2\k\,F_3+4B\,F_1^2\ ;&(A.9)\cr
			M^2_{KS}&=M_K^2+2\k\,F_1 +4B\,F_1F_3\ ;&(A.10)\cr
M_\s^2+M_{\s'}^2-2M_{\pi S}^2&=(4A+6B)\,F_3^2+(8A-6B)\,F_1^2-3\k\,F_3 \ ;&(A.11)\cr		
	(M_{\s'}^2-M_\s^2)\sin2\th_S&={\textstyle {2\sqrt2\over3}}(F_3-F_1)\big((4A+6B)\,F_3
	+(8A+6B)\,F_1+\k)\ ;&(A.12)\cr 
(M_{\s'}^2-M_\s^2)\cos2\th_S&= 
\ffract13\big((8A+6B)F_1^2 -(4A+6B)F_3^2+32AF_1F_3-\k(8F_1+F_3)\big).\qquad\ \ &(A.13)}$$

\newsect{References.}
\item{1.} M.~Gell-Mann, ``The Eightfold Way", Caltech report CTSL-20 (1961), unpub.
     Y.~Ne'eman, {\it Nucl.~Phys. \bf 26} (1961) 222; see also M.~Gell-Mann and
	Y.~Ne'eman, ``The Eightfold Way'', Benjamin, New York (1964).
 \item{2.} M.~Gell-Mann, {\it Physics Lett. \bf 8} (1964) 214; G.~Zweig, Erice Lectures 1964,
 in {\it Symmetries in Elementary Particle Physics}, A.~Zichichi, Ed. Academic Press,
 New York, London, 1965,  p 192, and CERN Report No. TH 401, 4R12 (1964) (unpublished).
\item{3.} A.~Zichichi, {\it Annals of Physics \bf 66} (1971) 405.
\item{4.} D.~Bollini et al, {\it Nuovo Cimento \bf 56A} (1968) 1173, {\it ibid., \bf 57A} (1968) 404;
A.~Zichichi, in {\it Evolution of Particle Physics} (Academic Press, New York, London, 1970), 299,
M.~Conversi Ed.
\item{5.} M.~Basile et al, Proceedings of the Int.~Conf.~on ``Meson Resonances and Related Electromagnetic
Phenomena", Bologna, Italy, 14-16 April 1971 (Editrice Compositori, Bolgna, 1972), 139, R.~Dalitz and
A.~Zichichi Eds.
\item{6.} D.~Bollini et al, {\it Nuovo Cimento \bf 58A} (1968) 289; {\it id.} {\it Phys.~Lett.
\bf 42B} (1972) 377.
\item{7.} The discovery of QCD went through various stages. Important papers were:\br
	O.W.~Greenberg, {\it Phys.~Rev.~Lett. \bf 13} (1964) 598; 
	M.Y.~Han and Y.~Nambu, {\it Phys.~Rev. \bf 139 B} (1965) 1006.\br
	H.~Fritzsch, M.~Gell-Mann and H.~Leutwyler, {\it Phys.~Lett.  \bf 47B} 
	(1973) 365. See also H.~Lipkin, {\it Phys.~Lett. \bf 45B} (1973) 267.\br
	 D.J.~Gross and F.~Wilczek, {\it Phys.~Rev.~Lett.  \bf 30} (1973) 1343; \br
	 H.D.~Politzer, {\it Phys.~Rev.~Lett. \bf 30} (1973) 1346,\br
	 and many more. See for instance G.~'t Hooft, in ``The glorious days of physics,
	Renormalisation of Gauge Theories", Erice lecture notes 1998, SPIN-1998/18, hep-th/9812203.
\item{8.} M.~Gell-Mann, {\it Physics  \bf 1} (1964) 63.
\item{9.} S.L.~Adler, {\it Phys.~Rev.~Lett. \bf 14} (1965) 1051; W.I.~Weisberger,
	{\it ibid. \bf 14} (1965) 1047; M.~Veltman, {\it Phys.~Rev.~Lett. \bf 17} (1966) 553.
 \item{10.} H.~Fritzsch, M.~Gell-Mann and H.~Leutwyler, {\it Phys.~Lett. \bf 47B} (1973) 365.
\item{11.} G.~'t Hooft, ``Gauge Theories  with  Unified  Weak,  Electromagnetic  and 
   Strong Interactions", in {\it E.P.S. Int.~Conf.  on  High  Energy  Physics}, 
   Palermo, 23-28 June 1975, Editrice Compositori, Bologna 1976, A.~Zichichi Ed.
\item{12.} G.~'t Hooft, {\it Phys.~Repts. \bf 142} (1986) 357.
   \item{13.} H.B.~Nielsen, ``An Almost Physical Interpretation of the Integrand of the 
     n-point Veneziano model", XV Int.  Conf.  on  High  Energy  Physics, 
     Kiev, USSR, 1970 ; D.B.~Fairlie and H.B.~Nielsen, {\it Nucl.~Phys. \bf B20}
	 (1970) 637;\br
	  Y.~Nambu,   Proc.~Int.~Conf. on {\it Symmetries and Quark models},
	 Wayne state Univ. (1969); Lectures at the Copenhagen Summer Symposium (1970).
 T.~Goto, {\it Progr.~Theor.~Phys. \bf 46} (1971) 1560;\br
 H.B.~Nielsen and L.~Susskind, CERN preprint TH 1230 (1970);
	L.~Susskind, {\it Nuovo Cimento \bf 69A} (1970) 457; {\it Phys.~Rev \bf 1} (1970) 1182.
\item{14.} H.B.~Nielsen and P.~Olesen, {\it Nucl.~Phys. \bf B61} (1973) 45;
 B.~Zumino, in {\it Renormalisation and Invariance in Quantum Field 
Theory}, NATO Adv.~Study Institute, Capri, 1973, Ed.~R.~Caianiello, Plenum (1974) p.~367.
\item{15.} G.~'t~Hooft, {\it Nucl.~Phys.  \bf B79} (1974) 276.
\item{16.} G.~'t~Hooft, {\it Phys.~Rev.~Lett.  \bf37} (1976) 8;
 {\it Phys.~Rev.  \bf  D14} (1976) 3432; Err.~{\it  Phys.~Rev. \bf  D18} (1978) 2199.
\item{17.} A.A.~Belavin, A.M.~Polyakov, A.S.~Schwartz and Y.S.~Tyupkin,  
	{\it Phys.~Lett. \bf 59 B} (1975) 85.
\item{18.} S.L.~Adler, {\it Phys.~Rev.}~{\bf  177} (1969) 2426.
\item{19.} J.S.~Bell and R.~Jackiw, {\it Nuovo Cimento \bf 60} (1969) 47. 
\item{20.} W.~Bardeen, {\it Phys.~Rev. \bf 184} (1969) 1848.
\item{21.}S.~Coleman, in {\it The Whys of Subnuclear Physics}, Erice July-August 1977,
A.~Zichichi, Ed. (Plenum Press, New York and London, 1979), p.~805.

\item{22.} E.~Witten, {\it Phys.~Lett.~B \ \bf 86B} (1979) 283.
\item{23.} G.~'t Hooft, {\it Nucl.~Phys. \bf B75} (1974) 461.
\item{24.} K.~Wilson, in {\it New Phenomena in Subnuclear Physics}, Erice 1975, Plenum Press,
New York and London, 1977, A.~Zichichi, Ed., p.~69.
\item{25.} S.~Weinberg, {\it Phys.~Rev. \bf D13} (1976) 974; {\bf D19} (1979) 1277; L.~Susskind,
{\it Phys.~Rev. \bf D20} (1979) 2619; S.~Dimopoulos and L.~Susskind, {\it Nucl.~Phys.
\bf B155} (1979) 237; E.~Eichten and K.~Lane, {\it Phys.~Lett. \bf 90B} (1980) 125.
\item{26.} G.~'t Hooft, in  "Recent Developments in Gauge Theories", Carg\`ese  1979, 
     ed. G.~'t Hooft et al., Plenum Press, New York, 1980,  
     Lecture III. Reprinted  in: G.~'t Hooft, "Under the Spell of the Gauge Principle", Adv
	 Series in Math. Phys. Vol {\bf 19}, World Scientific, Singapore, 1994, p. 352.
	 
\bye